# Hydrogen isotope separation using graphene-based membranes in liquid water

Xiangrui Zhang[a], Hequn Wang[c,d], Tiantian Xiao[a], Xiaoyi Chen[a], Wen Li[b,c], Yihan Xu[a], Jianlong Lin[a], Zhe Wang[d,e], Hailin Peng[b,c], Sheng Zhang[a] a  Key Laboratory for Green Chemical Technology of Ministry of Education, Haihe Laboratory of Sustainable Chemical Transformations, School of Chemical Engineering and Technology, Tianjin University, Tianjin 300072, China

b  Center for Nanochemistry, Beijing Science and Engineering Center for Nanocarbons, Beijing National Laboratory for Molecular Sciences, College of Chemistry and Molecular Engineering, Peking University, Beijing 100871, P. R. China

c  Beijing Graphene Institute, Beijing 100095, P. R. China.

d  School of Chemical Engineering & Advanced Institute of Materials Science, Changchun University of Technology, Changchun 130012, P. R. China.

e  School of Chemistry and Life Science, Changchun University of Technology, Changchun 130012, P. R. China.

**ABSTRACT:** Hydrogen isotope separation has been effectively achieved using gaseous $H_2/D_2$ filtered through graphene/Nafion composite membranes. Nevertheless, deuteron nearly does not exist in the form of gaseous $D_2$ in nature but in liquid water. Thus, it is a more feasible way to separate and enrich deuterium from water. Herein we have successfully transferred monolayer graphene to a rigid and porous polymer substrate PITEM (polyimide tracked film), which could avoid the swelling problem of the Nafion substrate, as well as keep the integrity of graphene. Meanwhile, defects in large area of CVD graphene could be successfully repaired by interfacial polymerization resulting in high separation factor. Moreover, a new model was proposed for the proton transport mechanism through monolayer graphene based on the kinetic isotope effect (KIE). In this model, graphene plays the significant role in the H/D separation process by completely breaking the O-H/O-D bond, which can maximize the KIE leading to prompted H/D separation performance. This work suggests a promising application of using monolayer graphene in industry and proposes a pronounced understanding of proton transport in graphene.

Large amount of heavy water is used as the neutron moderator in nuclear fusion reactors[1-2]. It is also widely used in laboratory spectroscopy and dynamics research, such as neutron scattering[3-5], isotope tracing[5-7], and solvent for proton nuclear magnetic resonance spectroscopy[8]. Due to its low abundance (~156 ppm) in nature, methods to gain heavy water with high purity is urgently required for industrial and research applications[9]. There are two traditional industrial methods to obtain heavy water: the Girdler-Sulfide process[10] and cryogenic distillation at 24 K[11]. However, both methods are relatively complicated, costly, and time-consuming due to their quiet low separation factor below 2. Therefore, a novel technology is essentially needed which could separate and enrich heavy water efficiently and economically.

Recently, graphene has been reported to be a feasible and effective solution for heavy water enrichment. With the aid of graphene, the separation ratio can reach as high as α~10 and energy consumption could be effectively reduced as low as 20 GJ/kg[12], which means it is a more economic way for heavy water enrichment. The pristine monolayer graphene can be visualized as a sieve that only allows protons and deuterons to pass through. Nevertheless, deuterons will transport more slowly due to its lower value of zero-point energy, which results in the separation of H and D. The monolayer graphene has a high protium-to-deuterium (H/D) separation factor of around 10, which was first demonstrated by Lozada-Hidalgo et al. in 2016 with micro-sized mechanical exfoliated single-layer graphene[13]. Subsequently, Zhang et al. measured the H/D separation ratio (~8) of macro-sized (~1 square inch) graphene fabricated by chemical vapor deposition (CVD) by mass spectrometry[12]. CVD is feasible to synthesize graphene with large size (even square meters) which is more useful for practical applications. However, defects will be inevitably introduced in graphene during the growth process[14-15], as well as the transfer process.[16] Unfortunately, these defects will deduce the H/D separation ratio. Thus, the defect-fixing technique is essentially required to achieve an excellent H/D separation ratio, especially for the large graphene membrane.

The H/D separation for the studies mentioned above was achieved using gaseous $H_2$ and $D_2$ in laboratory. In practical, it should be noted that deuteron does not exist in gaseous $D_2$ in nature but in liquid water, in the form of $D_2O$ or HDO. Thus, the separation and enrichment of deuterium from water is a more practical way compared to gaseous $H_2$ and $D_2$. Graphene needs to be transferred to a suitable substrate such as Nafion for this application. Nearly all previous studies employed Nafion as the substrate of graphene for gaseous H/D separation, which ascribes to the good transport ability of the channels in Nafion for protons and deuterons. However the technology they claimed encounters the stability issue of graphene devices in aqueous solution, due to the swelling of Nafion substrate, which will swell up to 15% after hydrated in wet state[17], and thus cause tearing and breakage problem inside the monolayer graphene.

Herein, in this work, the strategy to avoid swelling issue is to replace Nafion with a porous polyimide track-etched membrane (PITEM) with a swelling percentage of only 0.329% (100 RH%)[18]. The PITEM substrate can effectively prevent graphene from damage in liquid water. Furthermore, the strong adhesion between graphene and PITEM surface ensures the high integrity of graphene after transfer process and the long-term stability of the device[19]. Track-etch technique used on PITEM ensures the unhindered passage for protons with the help of artificial pores in the membrane. Moreover, in this work, interfacial polymerization (IP) was carried out to seal the defects and cracks created during the growth and transfer process to improve the integrity of the large-size graphene. As a result, the H/D separation ratio was markedly increased to 8.6 for IP sealed graphene membranes.

H/D separation by graphene was reported to originate from the zero-point energy difference (60 meV) between proton and deuterium. However, this theory could not explain the role of graphene in this process because the zero-point energy difference is only related to the vibrational frequencies of the O-H and O-D bonds, which does not involve any graphene's properties. To better understand the role of graphene in this separation process, a more specific model was proposed based on kinetic isotope effects (KIE) theory. Based on the new model, we predicted that H/D separation ratio for water electrolysis with and without graphene were 11.2 and 4.9, respectively, which were in agreement with the experimental data (8.6 and 4.0). This model implies the critical role of graphene in the isotopic separation process: completely break the O-H and O-D bonds to achieve the maximum KIE, as a result, high H/D separation ratio.

## RESULTS AND DISSCUSSION

**CVD Graphene on porous PITEM substrate.** Polyimide track-etched membrane (PITEM) is a type of nuclear track membrane with uniform pores and vertical channels (Figure 1), which is used in laboratory filtration, water filtration, cell culture growth, and environmental studies[20-22]. Polyimide also possesses good chemical stability and adhesion to graphene[19]. Here PITEM with hydrophilic surface is employed as a porous substrate to prevent the damage of graphene. Graphene grown on copper foil by CVD was transferred to PITEM using a polymethylmethacrylate(PMMA)-assisted transfer method[23], and the graphene/PITEM composite membrane was defined as PITEM-G. After transfer, graphene was characterized by Raman spectra, scanning electron microscope (SEM), and atomic force microscope (AFM) (Figure S1). The Raman spectrum showed a G/2D peak area ratio of 0.31, slightly higher than defect-free monolayer graphene (0.25), indicates the successful transfer of graphene24. The stability of PITEM-G in water was tested and compared with graphene/Nafion membrane (Figure S2). The graphene on PITEM-G still kept high integrity after 48 hours of immersion in water, while graphene/Nafion sample was full of cracks and defects due to the swelling of Nafion.

It can be seen from Figure 1(a,b) that the existing transfer process cannot guarantee that the graphene tightly adheres to the surface of PITEM in all regions, which is a negative factor for graphene integrity. Those defects formed during the transfer process have to be repaired to ensure the high H/D separation performance of graphene. The interfacial polymerization (IP)[23, 25-26] technique then was employed to seal defects and cracks of graphene. The monomers for IP are organic-soluble trichloromethane (TMC) and water-soluble m-phenylenediamine (MPD), which contacted and reactively polymerized at the water-organic interface[27-29]. The IP mechanism for defect repair is shown in Figure 1(c,d). Specifically, PITEM channels covered with defects have been fully blocked to ensure high separation selectivity. Since the intact graphene could prevent TMC in hexane from contacting MPD in water, no IP reaction occurred in the PITEM pores covered by undamaged graphene. In contrast, those two monomers for IP could penetrate and react at the crack regions, forming a polymer plug inside the PITEM channel. In this case, we defined this membrane as PITEM-G-IP.

PITEM (2.12% porosity, Figure 1e) without graphene was also treated with IP (PITEM-IP) to demonstrate that the IP reaction occurs selectively. It can be seen that all those native pores disappeared after IP treatment (Figure 1h), which proves the ability and effectivity of IP to plug the channels of PITEM. In contrast, the SEM of PITEM-G-IP (Figure 2g.) showed that the surface of the membrane still has a lot of channels left which has not been blocked by IP (1.20% porosity after repair), indicating that the IP reaction is only feasible within the defective pore channels, and eventually retains the pores covered by graphene for subsequent H/D separation.

Moreover, we designed a control experiment to prove that the polymers produced by IP were only generated in the channels of PITEM in the region where the graphene is defective, instead of all the PITEM's channels. (Figure S4) We employed oxygen plasma, which can effectively remove graphene from the composite membrane and keep the polymers untouched. It is easy to understand that the conductivity (measured by electrochemical impedance spectroscopy in liquid phase) of PITEM-G-IP is lower than that before IP because of the blockage of pores inside PITEM. After oxygen plasma treatment for PITEM-G-IP, the conductivity increased, which means that the channels covered with graphene were not plugged and exposed again after the graphene was removed. Therefore, it is a powerful proof to show that the IP only occurs in the channels of PITEM where they are not covered by graphene.

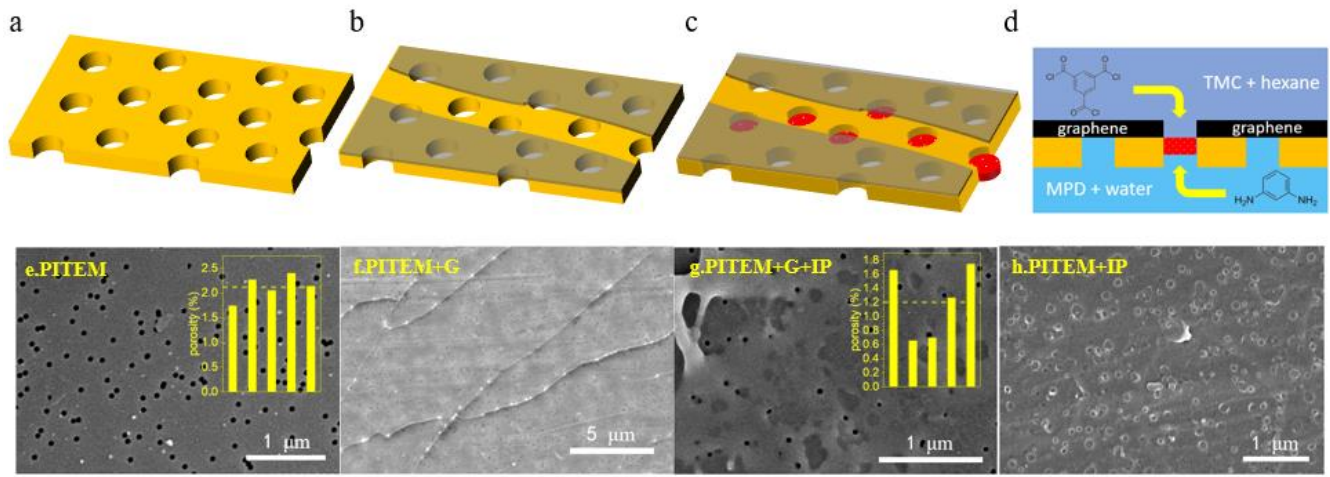

**Figure 1.** Graphene membrane fabrication and defect-sealing procedure. (a) PITEM without graphene. (b) monolayer graphene (the translucent gray layer) transferred to PITEM with small tears and defects (c) Sealed membrane (PITEM+G+IP) (d) Side view of the interfacial polymerization reaction, the yellow parts represent PITEM, and the red plug is the polymer produced by IP. (e) SEM image of blank PITEM (The upper left bar chart shows the porosity of five different membranes, the average porosity is 2.12%) (f) SEM image of PITEM+G, the white area is the tears and defects of graphene. (g) PITEM with graphene treated by IP, the pores under graphene defects are closed by polymer, and the reduction of porosity proves that some pores are sealed. (The upper left bar chart shows the porosity of five different membranes, the average porosity is 1.20%) (h) PITEM without graphene treated by IP, all the pores are sealed by polymers.

Then electrochemical impedance spectroscopy (EIS) was used to evaluate the ionic impedance of the membrane, which represents the flux of ion penetration through the membrane and thus determine whether the graphene repair is effective. We designed an experimental system shown in Figure 2a. The H-cell was filled with 0.5 mol/L $H_2SO_4$ as the electrolyte and connected to an electrochemical work station. The impedance of pristine PITEM (~50 mS/cm) was slightly lower than that of Nafion-XL[30], which indicates that PITEM could hardly block the migration of hydrogen ions when used in a liquid phase. After IP for PITEM-G, the resistance of the membrane was dramatically increased by 4-fold. Since polymer plugs were selectively produced in the pores, the increment of impedance value proves that IP repair can prevent the leakage of hydrogen ions passing through the defects.

Furthermore, the conductivity of graphene was also used to evaluate the integrity of graphene. The lower the conductivity, the higher the integrity of graphene. To show the high integrity of graphene sealed by IP, a comparison of the hydrogen ion conductivity for graphene in this work and some literature results can be seen in Figure 2c. The hydrogen ion conductivity of the repaired graphene was estimated to be 39.1 mS/cm$^2$ (effective area 1.77 cm2), which is comparable to the value of graphene in some micro-size devices(4 mS/cm$^2$, 1.96×10-7cm$^2$, loaded on a silicon chip) and is one to two orders of magnitude lower than the values reported in other graphene devices with defects and cracks[31-36] (With the similar size, the lowest conductivity reported in literatures is 1667 mS/cm$^2$, much higher than our work). This result suggests that IP sealing is very effective to repair defects and cracks on large area of graphene and achieve highly efficient H/D separation.

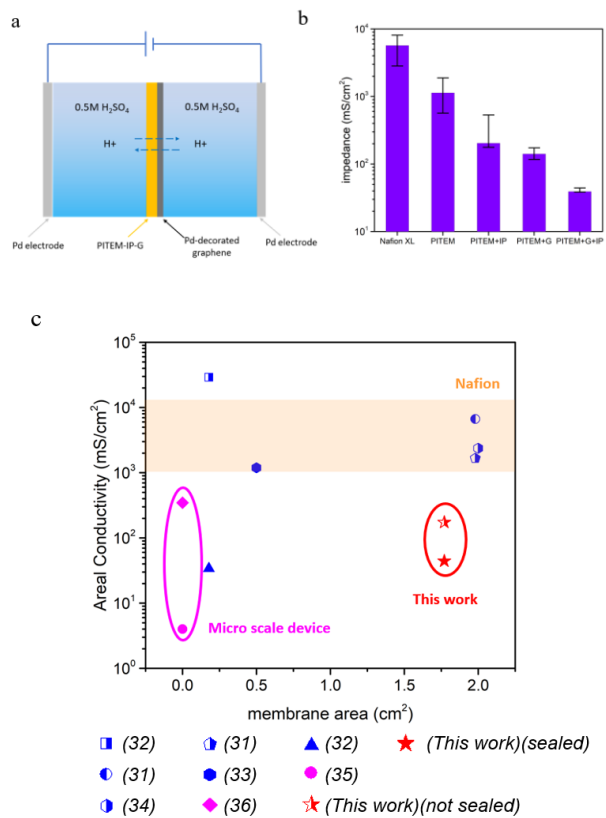

**Figure 2.** (a) Schematic diagram of hydrogen ion impedance test device (b) bar chart of hydrogen ion impedance. Nafion XL and blank PITEM were used as control groups, the impedance decreased significantly after loading graphene and carrying on IP. (c) Experimentally measured areal proton conductivity values for graphene in this work and literature. The conductivity of Nafion was collected from literature.

**H/D Separation in liquid water using graphene.** The H/D separation ratio for graphene devices can be measured by mass spectrometry. As shown in Figure 3(a), a semi-solid reactor was used to obtain the accurate separation ratio. A platinum mesh electrode was used as the anode. The electrolyte was sulfuric acid solution (pH=0) with H/D=1:1. The graphene device (Figure 3b) acted as the cathode, and palladium was decorated on the surface of graphene to ensure that the protons/deuterons are reduced after passing through the graphene. Once bias applied, H$_2$, D$_2$ and HD gas diffused from the backside of the graphene device into a sealed chamber. Subsequently, the H/D separation ratio(α) was calculated using the equations below, where [H$_2$][D$_2$][HD] represent the concentrations of hydrogen, deuterium and hydrogen-deuterium gas, respectively

$$\alpha = \left(\frac{1}{2}[HD] + [H_2]\right) / \left(\frac{1}{2}[HD] + [D_2]\right)$$

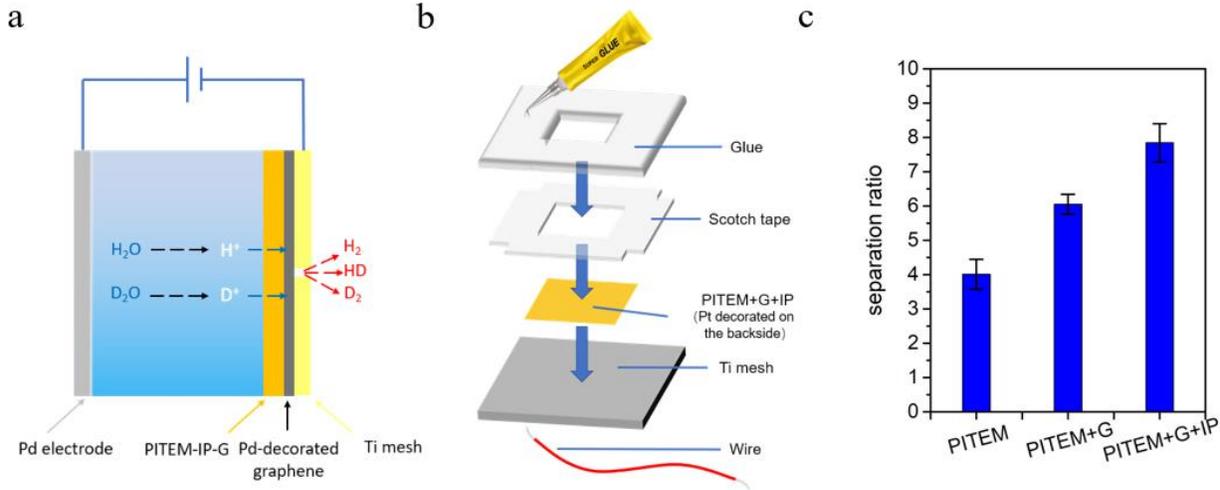

**Figure 3.** (a) Schematic diagram of the test apparatus for hydrogen-deuterium separation ratio. H-cell was used to set up the test device, and the electrolyte was the mixture of 0.25 mol/L H$_2$SO$_4$ and 0.25 mol/L D$_2$SO$_4$ (b). Schematic of the graphene device, which was placed on the cathode side of the H-cell. (c)The separation ratio of PITEM, PITEM-G, and PITEM-G-IP.

The H/D separation factor was 4.0±0.4 for pristine PITEM, and it increased to 6.0±0.3 for PITEM-G. The former selectivity was the result of H/D isotopic effect of hydrogen evolution reactions over Pt catalysts. Graphene in PITEM-G exhibited a much lower H/D separation factor in aqueous solution than that (>8) in gaseous H$_2$/D$_2$ condition[12]. This poor performance could be attributed to large defects or cracks of graphene created in liquid water, and thus a dramatic decrease for separation ratio. In contrast, the separation ratio was increased as high as 8.6 for PITEM-G-IP in aqueous solution, suggesting that interfacial polymerization selectively enabled defects sealing in graphene.

It is crucial to elucidate the fundamental mechanism for the hydrogen-deuterium separation through monolayer graphene. In recent years, Theoretical studies have proposed different mechanisms (via hydrogenation[37-38] or protonation[39]) and transport pathways (straight perpendicular path through the center of the hexagonal ring[13] or via defects such as Stone-Wales(5577)[40-41])[42]. However, those explanations were not satisfactory so far. For instance, Hu et al.[43] claimed that proton transport across graphene is a thermally activated process, which could be demonstrated using the Arrhenius equation (Eq. 1). Lozada et al.[13] paved further and they proposed that the H/D separation ascribes to the difference of zero-point energy (60 meV) between hydrogen and deuterium which contributed a high H/D separation ratio, and the H/D separation ratio (α) can be derived from the equation(Eq. 2). where ΔE=E$_D$-E$_H$=60 meV (E$_D$ and E$_H$ represent the energy barriers required for protons and deuterons to pass graphene, respectively).

$$k_H = A e^{-\frac{E_H}{RT}} \quad (1)$$

$$\alpha = k_H/k_D = exp(\frac{\Delta E}{RT}) \quad (2)$$

Although those two theories were consistent with the experimental data, the explanations focus only on the initial state of the separation process, whereas ignore the role of graphene in the H/D separation process. For water electrolysis process catalyzed by Pt (without graphene), ΔE is only related to the initial state, the separation ratio (α∼10) can also be calculated by the theories above. However, the experimental measured H/D separation ratio for Pt-catalyzed water electrolysis is only about 2-6 [44-46], which was much lower than the calculated value.

This large deviation between theoretical and experimental value may come from the ignoration of the transition states in the separation process. Atoms have zero-point energies only in boundary states (e.g., chemical bonding is a boundary force). As for the H/D separation process, the chemical bonds (O-H, O-D) are not completely broken when the H/D atoms are in the transition state (O-H-Pt, O-D-Pt), and the zero-point energy still exists. Nonetheless, these ZPE in transition state are not involved in the model of Lozada et al. That is, H and D are considered that they have the same energy barrier in transition state. In fact, the energy barriers in transition state of H and D are still different due to the ZPE, which means that the above theories are not applicable in this work.

**A new model proposed for proton transport through monolayer graphene.** We proposed a theoretical model based on kinetic isotope effects (KIE) for the proton transport through graphene in this work, which takes consideration of the ZPE of the transition state. Kinetic isotopic effect (KIE) refers to the phenomenon that isotopes react on the catalyst surface with different rates [47]. It is estimated that the ratio of the H/D reaction rate equals to the H/D separation ratio in this work. To simplify the transfer process of protons and deuterons, we assumed that the energy barrier is the same for protons and deuterons excluding ZPE, as shown in Figures 4a and 4c. This assumption is reasonable and reliable because the electron Schrodinger equation does not involve the mass of the nucleus. In short, the electromagnetic forces to protons and deuterons are identical because they have the same amount of charge, and thus the same energy barrier.

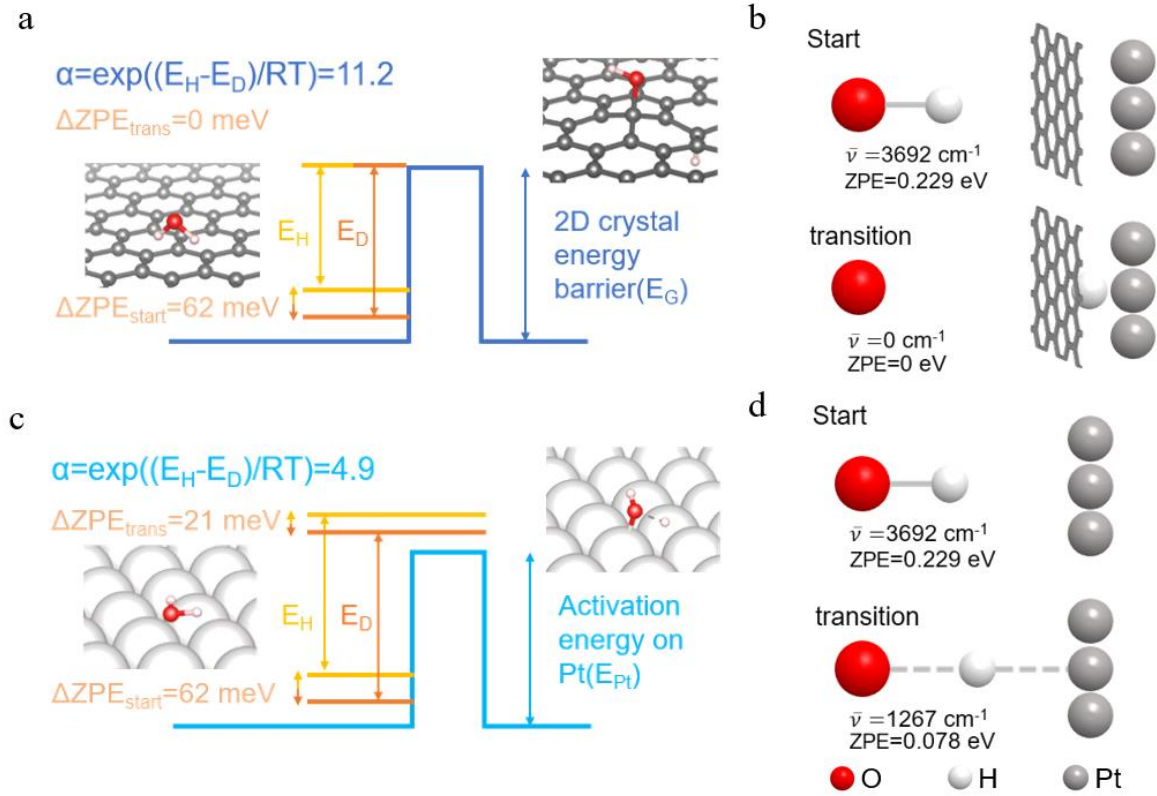

**Figure 4**. schematic of model based on kinetic isotopic effect (KIE) for proton and deuterium transport. (a)The dark blue line represents the 2D crystal energy barrier graphene. The orange line represents the zero-point energy of hydrogen and deuterium in different states. Inset (from left to right, respectively): (i)Proton adsorbs on the graphene surface. (ii) Proton completely traverses the graphene. (b)The change of ZPE of protons and wave number of O-H bonds during the process of protons passing through the graphene (c) The light blue line represents the reaction energy barrier of HER on Pt. Inset (from left to right, respectively): (i) Proton adsorbs on the Pt surface. The orange line also represents the zero-point energy of hydrogen and deuterium in different states. (ii) Proton forms transition states at the Pt surface. (d) The change of ZPE of protons and wave number of O-H bonds during the process of protons react with Pt.

Based on the assumption above, the H/D separation ratio is proportional to the difference in zero-point energy of the initial state and transition state. In this work, H/D separation by water electrolysis at the surface of Pt(111) without (Figures 4a and 4b, case 2) or with monolayer graphene (Figures 4c and 4d, case 1) were discussed to illustrate the rationality of the new model. The measuring conditions were the same for both without/with cases. The initial state is the water molecules absorbed on graphene or Pt(111), and the ZPEH-start (0.229 eV) as well as ZPED-start (0.167 eV) can be calculated based on equations 3 and 4[47]. Where h is the Planck constant, ν is the vibration frequency, $\bar{v}$ is the wave number (obtained by DFT), c is the speed of light, k is the chemical bond force constant (The value of k is the same for both H and D), $m_1$ is the atomic mass ratio of hydrogen/deuterium, and $m_2$ is the atomic mass of oxygen.

$$E = \frac{1}{2}hv = \frac{1}{2}h\bar{v}c \quad (3)$$

$$v = \frac{1}{2\pi}\sqrt{\frac{k}{m_r}} \quad m_r = \frac{m_1 m_2}{m_1+m_2} \quad (4)$$

$$E_H = E_{Pt} - ZPE_{H-start} + ZPE_{H-tran} \quad (5)$$

$$E_D = E_{Pt} - ZPE_{D-start} + ZPE_{D-tran} \quad (6)$$

$$\Delta E = E_H - E_D \quad (7)$$

We first discuss the water electrolysis at the surface of Pt. In this case, the protons and deuterons react directly on the surface of Pt(111), the O-H and O-D bonds will not break completely at this step because of the transition state(Figure 4d). So, when the reaction occurs, H/D atoms need to conquer the origin energy barrier ($E_{Pt}$) and the zero-point energy ($ZPE_{H-trans}$, $ZPE_{D-trans}$). In other words, the zero-point energy of the initial state is not fully exploited without graphene because of the ZPE of the transition state. According to Equations 2,3,4,5,6,7, the ΔE is 41 meV compared to 60 meV calculated by Lozada's theory, which resulted in a decrease in the H/D separation ratio from 11.2 to 4.9. The separation ratio of 4.9 is close to the value reported[44-46](α=2∼6) as well as our experiment data(α∼4.0±0.4) (Figure 4c).

The membrane with graphene prevents the direct contact of protons and deuterons with the catalyst, so the H and D atoms need to be separated from the O atoms to pass through the monolayer graphene and reach the surface of Pt to trigger the hydrogen evolution reaction (HER) (as shown in figure 4b). In this process, the O-H/O-D vibration frequency decreased to zero due to the completely broken O-H and O-D bonds. According to equations 2,3,4,5,6,7, the $ZPE_{H-trans}=ZPE_{D-trans}=0$, and the separation ratio is 11.2. This result also agrees with the value reported in literature (α∼10) and our experiment data(α∼8±0.6), which illustrates the model's rationality in a broader range of scenario comparing with previous theory (all the data are collected in table 1).

**Table 1. comparison of experiment value and calculated value**

|  | O-H wave number/cm$^{-1}$ | O-D wave number/cm$^{-1}$ | $ZPE_{H-start}$/eV | $ZPE_{D-start}$/eV |
|---|---|---|---|---|
| Experiment | 3373 | 2492 | - | - |
| Calculation | 3692 | 2686 | 0.229 | 0.167 |
| Literature | 3400 | 2500 | 0.20 | 0.14 |

|  | ZPEH-trans/eV | ZPED-trans/eV | Separation ratio (with G) | Separation ratio (without G) |
|---|---|---|---|---|
| Experiment | - | - | 8.0 | 4.0 |
| Calculation | 0.078 | 0.057 | 11.22 | 4.9 |
| Literature | - | - | 10 | 2∼6 |

The separation ratio of 11.2 was estimated by our new model, which was close to 10 calculated by Lozada's model. The difference between the two values may be attributed to the error of DFT calculation. According to Eq2, the separation ratio α only relates to the ΔE. In the case with graphene, $\Delta E=\Delta ZPE_{start}-\Delta ZPE_{trans}$ (Figure 4a), whereas, $\Delta E=\Delta ZPE_{start}$ in Lozada's model, which the $\Delta ZPE_{trans}$ was not involved., since the broken bond brought by graphene happens to make $\Delta ZPE_{trans}=0$ the correct value could be given by Lozada's model. However, Lozada's model is still limited for explaining all transport mechanism.

As mentioned above, the $\Delta E=\Delta ZPE_{start}-\Delta ZPE_{trans}$, and the $\Delta ZPE_{start}$ is the same (62 meV) for both without- and with- case, so it is the difference of $\Delta ZPE_{trans}$ (0 meV for the case without graphene, 21 meV for the case with graphene), resulting in the profound change of separation ratio. Thus, it can be assumed that the transition state plays a decisive role in separating hydrogen and deuterium. The existence of graphene gives rise to a completely different transition state, which could effectively increase the separation ratio from 4.9 to 11.2. In order to cross the monolayer graphene and access the catalyst, the H/D atoms are forced to completely separate from the O atoms (completely broken of O-H and O-D bonds). In other words, the maximum of KIE could be achieved with the aid of graphene. Meanwhile, it is reported that the KIE could reach its maximum value when the chemical bonds are broken completely [47], which proves the rationality of our theory.

Comparing with the model proposed by Lozada in 2016, our new model considers the effect of the transition state as well as clarifies the role of graphene in the H/D separation process. Furthermore, the new model is more applicable to describe the H/D separation process in different conditions (catalysis, with or without graphene, electrolytes, pH and so on). Based on this model, we would like to suggest some perspectives: (i) 2D materials including but not limited to graphene, can help reach the maximum H/D separation ratio; (ii) various chemical conditions, which could influence the ZPE, will affect the maximum H/D separation rate; (iii) the strategy is equally effective for H/T and D/T or even isotopes of other elements. All in all, separating hydrogen isotopes by monolayer graphene-based membrane to get the maximum separation ratio is a promising and effective strategy which has the potential to be widely applied.

## CONCLUSION

In conclusion, we employed PITEM as the substrate for graphene to avoid the swelling issues of Nafion substrates in liquid water. Meanwhile, the defects in large-area monolayer graphene were successfully sealed by interfacial polymerization (IP) technology. After sealing, the H/D separation ratio was improved from 6.0 to 8.6, demonstrating the feasibility of IP repair for graphene breakage. Moreover, this work developed a new model based on kinetic isotope effects (KIE). Graphene prevents the direct contact of hydrogen and deuterium atoms with the catalyst, so that hydrogen and deuterium atoms can only pass through graphene when the O-H and O-D bonds are completely broken. As a result, the KIE value was enhanced close to the theoretical maximum value (~10). The H/D separation ratio calculated by the model is 11.2 and 4.9 for the presence (α~11.2) and absence (α~4.9) of graphene, respectively. These theoretical values are close to the experimental value (α~8.6 and 4.0), indicating the rationality and the universality of our new model.

## METHODS

**Membrane design.** We designed and fabricated a composite membrane with graphene on top surface of a porous polymer substrate. Hydrophilic polyimide track-etched membranes (PITEM) (it4ip S.A.) were used as the porous support layer, due to the well-tailored cylindrical cores with optimized diameters which can provide good mechanical support for graphene and allow ions to pass through at the same time. It also shows good chemical stability and adhesion to graphene.[19] Graphene used in this work was grown by chemical vapor decomposition (CVD) on Cu foil, and transferred to PITEM membrane via a poly (methyl methacrylate)- (PMMA)- assisted process. Then we carried out interfacial polymerization (IP) treatment for the composite membrane to repair the defects existed in graphene which were created in the process of graphene synthesis and transfer process. Finally, the repaired graphene-based membrane was placed into an electrochemical pump which was connected with mass spectrometry to detect the separation performance of hydrogen and deuterium.

**Synthesis of graphene.** Graphene growth was conducted under $H_2$ (99.999% purity, the concentration of $H_2O$ ≤ 3 ppm), Ar (99.999% purity, the concentration of $H_2O$ ≤ 4 ppm), $CH_4$ (99.999% purity, the concentration of $H_2O$ ≤ 3 ppm), and $O_2$ (0.1% diluted in Ar, the concentration of $H_2O$ ≤ 3 ppm). After the growth procedure, the samples were cooled to 700 °C under the same atmosphere as growth and then were moved outside of the hot zone under $H_2$ (6-in. tube furnace) or $N_2$ (A3-size industrial system), according to the literature[48]

**The fabrication of graphene/PITEM membrane.** Graphene was normally grown on both top and back sides of copper foils, and the top side basically shows higher quality. Therefore, top side was spin-coated with PMMA and then baked at 130 °C for 3 min. Subsequently, the back side of copper foil was exposed to air plasma for 3 min to remove the residual graphene. Then, the copper foil was etched with 1 M Na2S2O8 solution by floating the whole PMMA/graphene/copper foil on the surface of the solution. Then the free-standing PMMA/ graphene membrane was washed with deionized water and then transferred onto PITEM. After air drying, the PMMA/graphene/PITEM was immersed in acetone at 80 °C for about 10 min to remove PMMA. At this point, graphene was successfully transferred to PITEM.

**Interfacial polymerization on graphene.** Defects and tears in graphene were selectively sealed by interfacial polymerization (IP). IP is a reaction that takes place at the interface of two monomers, the reactants in different phases would react when they contact with each other. The reactants can contact and react through the defects and tears inside monolayer graphene, but they cannot pass through graphene. As a consequence, the product of IP would seal the defects and tears in graphene effectively. We carried out IP process by using an unstirred 7ml Franz cell with a 20mm orifice to. Graphene-based membrane was placed at the middle place of the cell which kept the graphene side down and PITEM side up. After clamping the clip on the cell, filled the upper chamber with 0.05wt% Trimesoyl chloride (TMC) in hexane and wait for 10 minutes to allow TMC to fully diffuse in PITEM. Then filled the lower chamber with aqueous 0.05wt% m-Phenylenediamine(MPD) and 0.2wt% Sodium dodecyl sulfate(SDS) slowly without any bubbles and made the reactants to react for 30 minutes. Then the reaction was stopped by pipetting out the TMC on the upper chamber, then the graphene-based membrane was put into the oven and dried at 100 °C for 10 min to ensure cross-linking effect. Finally, the membrane was immersed in deionized water to remove the residual reactants.

**O-plasma treatment of graphene device.** To ensure the polymeric plugs formed selectively at the defect regions, we carried out a control experiment to measure the impedance of the membrane before and after O-plasma treatment, which was implemented by plasma equipment (CTP-2000 K, Suman, China) (power of ~4.95 W, oxygen flow rate of ~30 sccm and maintain for 1 min).

**Electrochemical impedance spectroscopy test.** We used electrochemical impedance spectroscopy (EIS) to test the impedance of the repaired graphene-based membrane in acid solution. EIS was carried out in a two-chamber electrochemical cell(H-cell) with two electrodes. The graphene-based membrane was clipped between the two electrode chambers. The two Pt plate electrodes, serving as working and counter electrodes, were used to supply electrical power for the whole system. 0.5 mol/L $H_2SO_4$ was used as the electrolyte.

EIS experiment was conducted using an electrochemical workstation (Zahner Zennium E4), which is in potentiostatic mode with a frequency range of alternating current from 100 Hz to 1MHz. The data points were collected on the logarithmic scale. The AC amplitude was set as 20 mV. EIS data were fitted by a nonlinear least-squares algorithm with the aid of Zview 3.0 software package.

**Preparation method of the electrochemical pump (graphene device).** The membranes (Nafion, PITEM, PITEM-G, PITEM-G-IP, prepared by the method mentioned above) were cut into the size of 1.2 cm×1.2 cm. Subsequently, Pt ions sputtering was performed on the graphene side by the ion sputtering instrument(108AUTO). The sputtering time was 20 seconds. Then the membrane was fixed on the center of the titanium mesh (circle of 2 cm diameter, used as the current collector and support) by using scotch tape. To ensure that the device has good conductivity and leak-proof, we used the glue (Araldite) to seal the top of the device. The glue could prevent the electrolyte from contacting the Ti mesh, avoiding side-effects such as water electrolysis. Moreover, it could improve the mechanical strength of the device and prevent the damage of graphene during the testing operation. After 24 hours, when the glue is totally air dried, the wire was connected with Ti mesh by using conductive silver paste. Then another 24 hours was needed to air dry the silver paste.

**Mass spectrometry measurements.** The H-cell was used as a heavy water separation reactor. PITEM-G-IP device was sandwiched between two chambers of H-cell, and one chamber was filled with electrolyte (0.5 mol/L $H_2SO_4$, 0.5 mol/L $D_2SO_4$ dissolved in 1:1 mixture of water and heavy water). The anode was 1*1 cm Pt mesh electrode and the cathode was connected with the wire of the PITEM-G-IP device. Another chamber was hermetically sealed, which served as a gas collector connecting with a sealed gas bag.

A helium leak detector (Leybold Phoenix Quadro) was used to test the separation effect of repaired graphene-based membrane. The key part of the instrument was a sealed pipeline with flange and mass spectrometry. A device with tiny leakage holes was clipped between the flanges to ensure that the gas in the upper chamber could transport to the lower chamber at a relative small flow rate. The leakage hole can ensure that the lower chamber had a high vacuum so that the molecular pump of the mass spectrometer can operate normally. The gas in the upper chamber comes from the reaction gas generated by the heavy water separation device collected in the sealed gas bag. The gas flow rate from the sealed bag in the upper chamber was controlled by a mass flow meter (Horiba Stec S48 32). The small amount of mixture gas generated by passing through graphene-based membrane was pumped into the mass spectrum. As The amount of gas in the lower chamber was much less than that in the feed chamber, the hydrogen/deuterium atomic ratio in the feed chamber can be considered 1:1. Then the following formula can be used to describe the hydrogen-deuterium separation ratio.

$$\alpha = \left(\frac{1}{2}[HD] + [H_2]\right) / \left(\frac{1}{2}[HD] + [D_2]\right)$$

**Methods of Density Functional Theory (DFT) calculations.** All density functional theories (DFT) were performed with Vienna Ab Initio Simulation Package (VASP)[49], using the projector augmented wave (PAW) pseudopotentials containing gradient-corrected Perdew-Burke-Ernzerhof exchange correlation[50-51]. The valence electronic states were expanded in a plane wave basis set with a kinetic cutoff energy of 400 eV. In the geometry optimizations, self-consistent field computations were repeated until the sum of forces acting on the relaxed atoms was below 0.05 eV Å-1. A vacuum thickness of 15 Å between the slabs was applied perpendicular to the surface to prevent spurious interactions between the repeated slabs. The climbing image general nudged elastic band (CI-NEB)[52] method was performed between the initial and final states to determine the transition states. Its results were further confirmed using frequency analysis. The metal surface was studied as slabs, for Pt(111), four layers with p(4×4) super cell were used. The top two layers were relaxed, whereas the bottom two layers were kept unchanged. Brillouin zone sampling was carried out by using 1×1×1 Monkhorst-Pack grids, which can suffice for well-converged results. The graphene was represented by a single layer slab with a 4×4 super cell. A 3×3×1 Monkhorst-Pack k-point mesh was used. In addition, the effect of van der Waals forces was involved by considering the weak interaction[53].

## ASSOCIATED CONTENT

**Supporting Information**.

## AUTHOR INFORMATION

**Corresponding Author**

* sheng.zhang@tju.edu.cn
* ntuliwen@outlook.com

**Author Contributions**

Xiangrui Zhang and Hequn Wang. contributed equally to this paper.

**Funding Sources**

Any funds used to support the research of the manuscript should be placed here (per journal style).

**Notes**

Any additional relevant notes should be placed here.

## ACKNOWLEDGMENT

The authors are grateful to the financial support from the National Nature Science Foundation of China (Grant No. 22078232 and 21938008), and the Science and Technology Major Project of Tianjin (Grant No. 20JCYBJC00870 and 19ZXNCGX00030).

# Supplementary notes

## Graphene transferred to PITEM

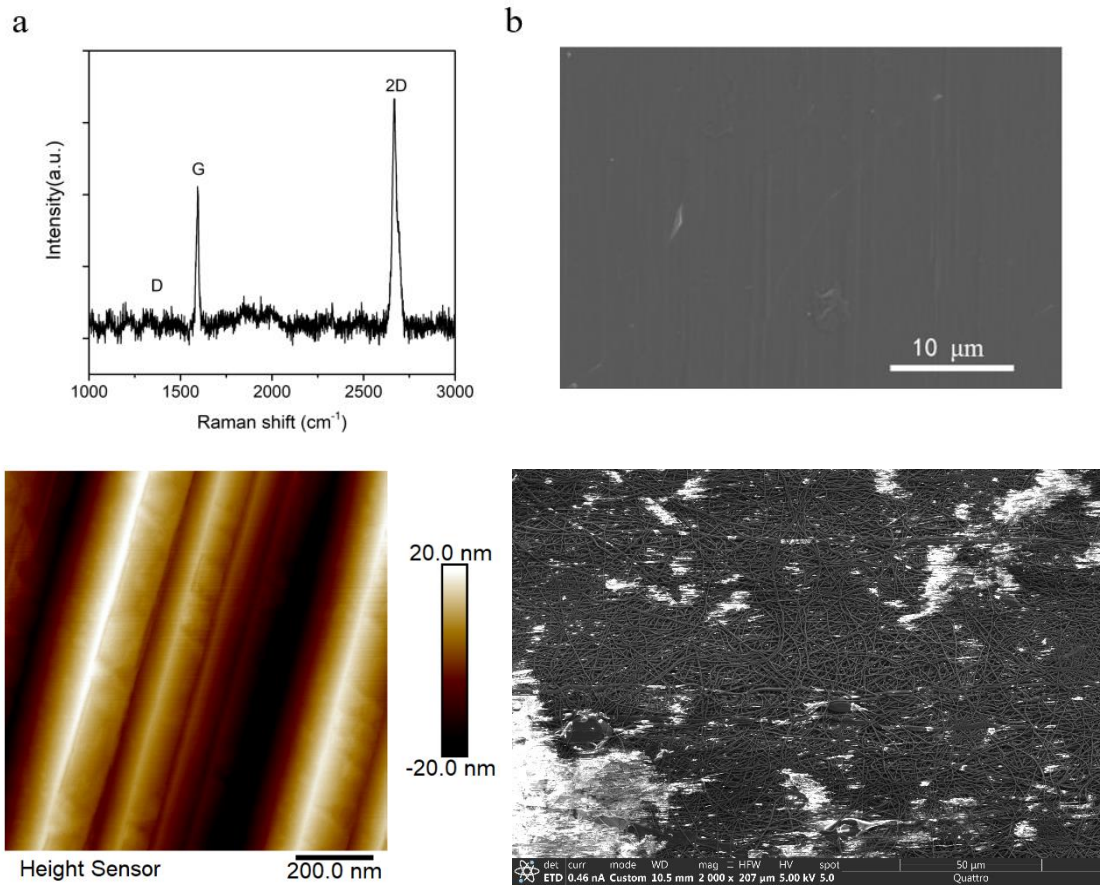

Figure S1. (a)Raman spectra of monolayer graphene on Cu foil (prepared by CVD). (b) SEM image of monolayer graphene transferred onto PITEM without Pt ions sputtering (c) AFM image of monolayer graphene on Cu foil. The larger depth change is caused by the folding of Cu while the smaller depth change is caused by the wrinkle of graphene. (d)SEM image of broken graphene, the white area represents the defects on graphene

## Stability in liquid phase

To prove the PITEM+G device has good stability in the liquid phase, we immersed the PITEM+G and Nafion+G into water for 48h. The Nafion + G device was observed to produce significant curling while the PITEM+G did not. It can be expected that the breakage of graphene is because of the swelling of Nafion in water. Then take samples for observation by SEM.

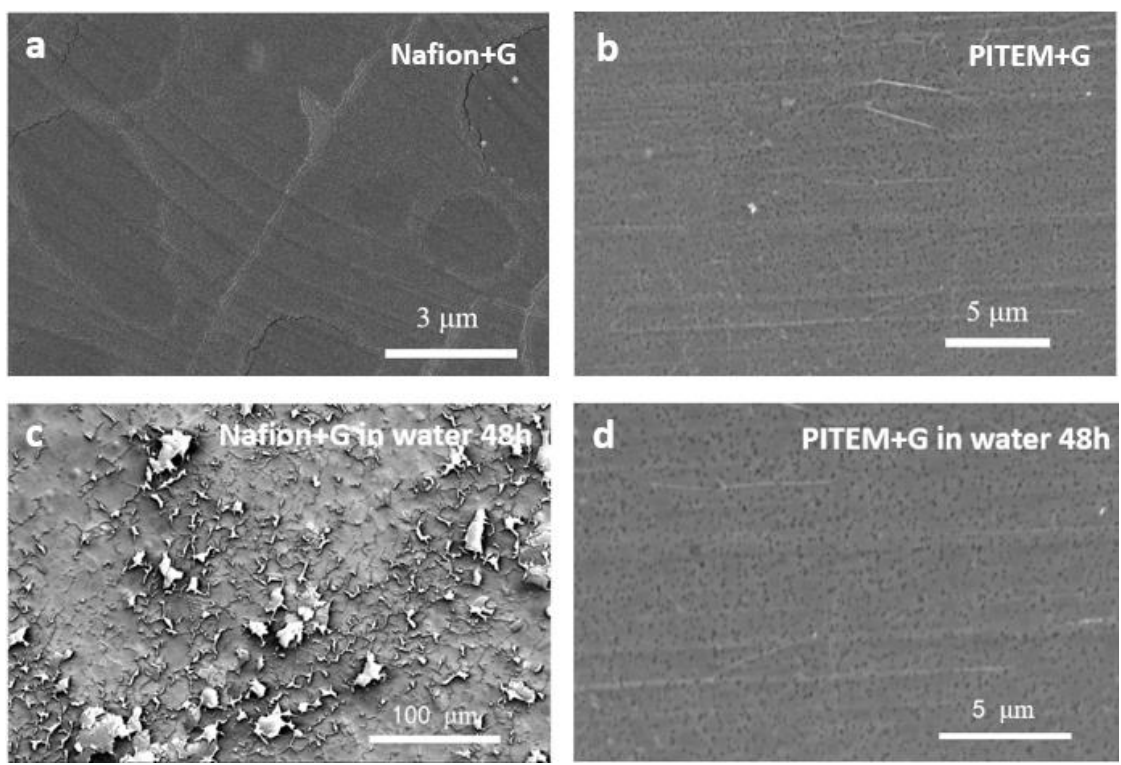

Fig S2. SEM image of monolayer graphene supported on Nafion and PITEM before(a,b) and after(c,d) immersed in water for 48h

## Effect of interfacial polymerization on porosity

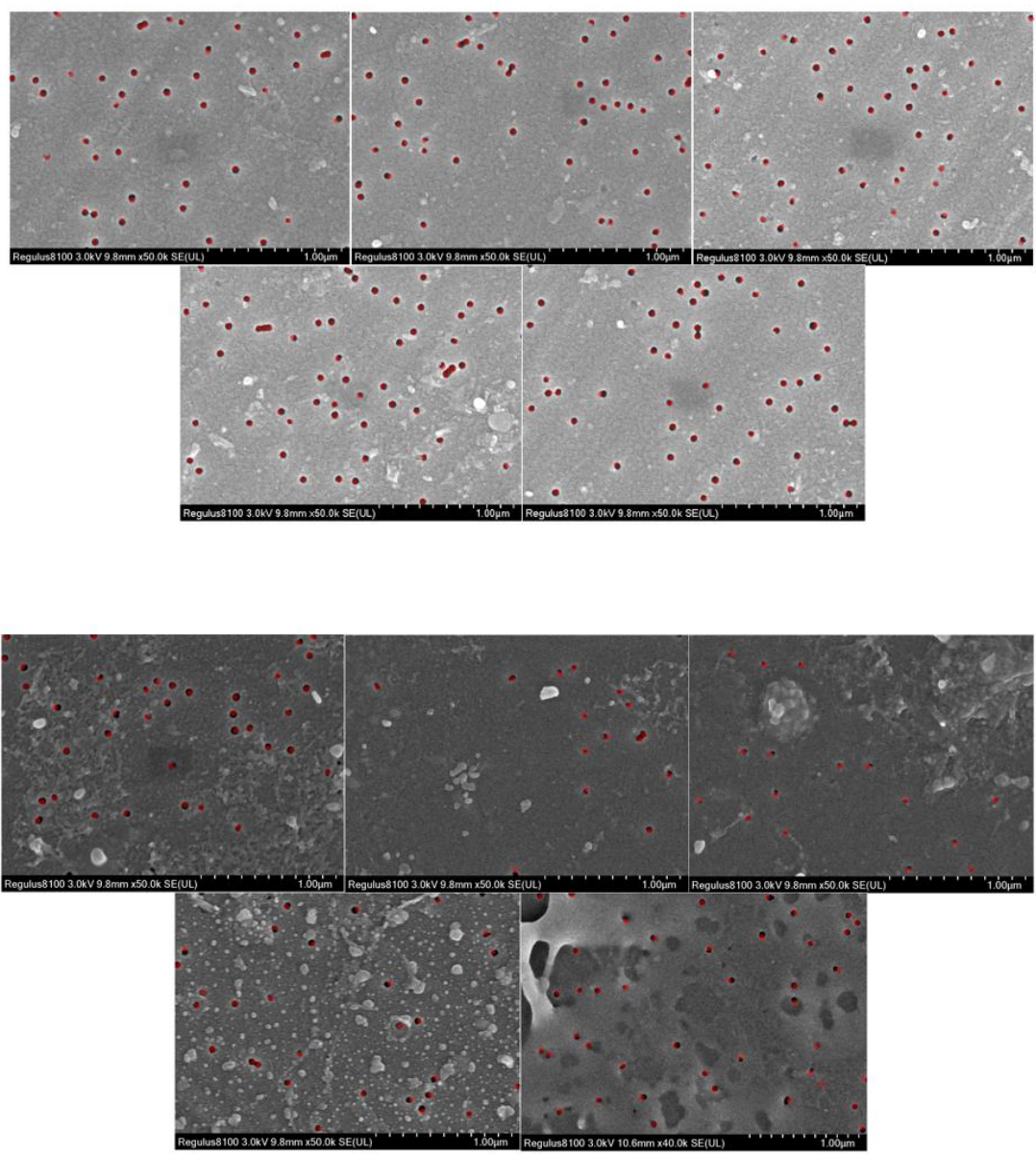

Fig S3. Statistics of porosity of PITEM before IP (upper) and after IP (lower). SEM figures are taken from different areas of the membrane to ensure the reliability of data. The red dot represents the hole, calculate the porosity (Fig 1) by averaging the number of holes.

## IP selectivity occurs in PITEM's channel

In addition, we design a control experiment to prove the polymers produced by IP only generated in the channels of PITEM in the region where the graphene is defective, instead of all the PITEM's channels. Firstly, we carried out O-plasma treatment, which can remove the graphene transferred on PITEM, on PITEM+G+IP (experimental group), then we use electrochemical impedance spectroscopy (EIS) to test the membrane. In this way, the ion impedance of the membrane can be obtained. In addition, we used the same experimental conditions to carry out O-plasma treatment on PITEM+IP(control group), then measured the ion impedance by EIS. (Fig.S4a) The ion impedance of the experimental group recovered to the level before graphene transfer after O-plasma treatment, while the control group did not. The ion impedance of the control group did not recover, indicating that the polymer plug could not be removed by O-plasma treatment. But the ion impedance of the experimental group had recovered to the level before graphene transfer, which shows that the polymer plugs are only generated in a few areas, and combined with data in Fig 3, it can be proved that these areas are the defects of graphene. As a consequence, we can safely draw a conclusion that the polymer plugs selectively grow in the channels of PITEM at the defects and tears of graphene.

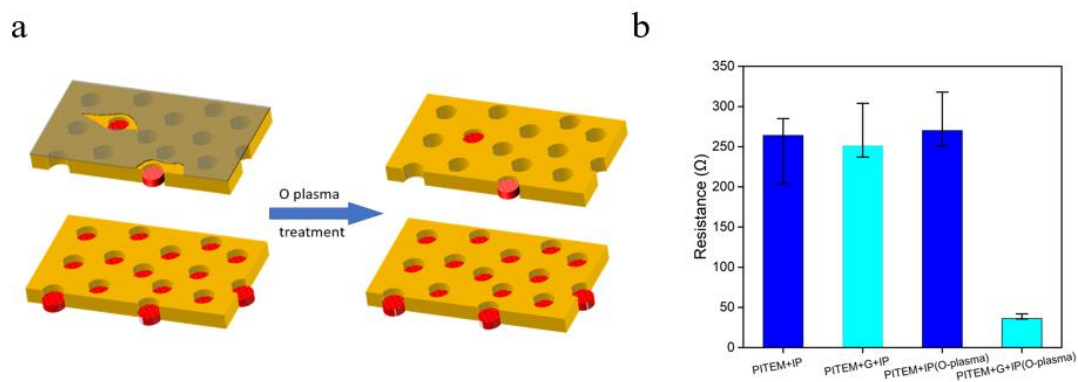

Fig S4. (a)procedure of control experiment. O plasma treatment was used for (I)PITEM+G(upper left) (II)PITEM(bottom left) (b)resistance of membrane, the error bar are taken from the maximum and minimum values of multiple groups of experimental values

## Optimization of interfacial polymerization conditions

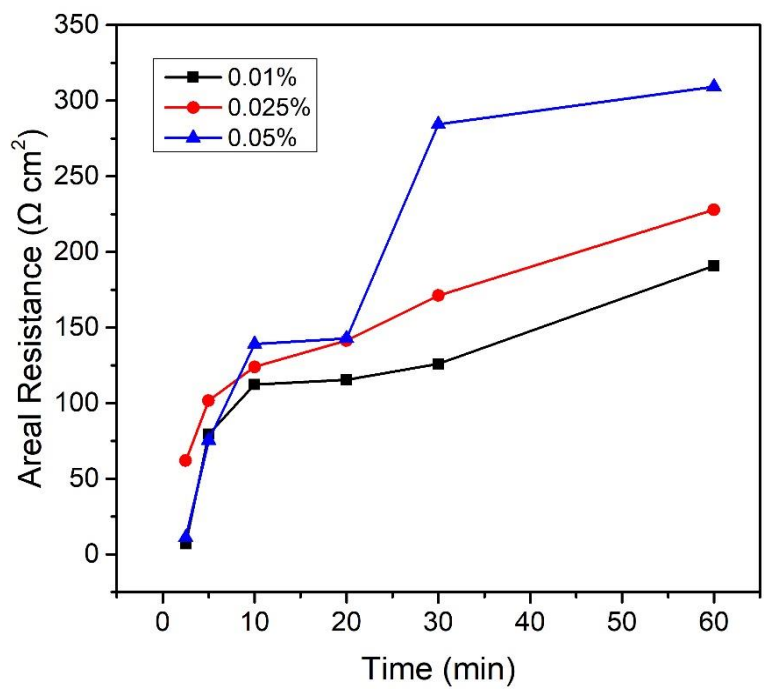

Fig S5. Optimization of interfacial polymerization conditions

# Calibration curve of helium ophthalmoscope

To ensure high accuracy of helium ophthalmoscope, we measured entry and exit gas composition differences on the helium ophthalmoscope. Firstly, we feed the feed chamber with a known ratio of $H_2/D_2$ gas, which was adjusted by the flowmeter. Then we measured the H/D ratio of the gas whose H/D ratio is known. Subsequent fit the above data by matlab, we get the following equation. Thus, the correction factor of the helium leak detector is 1.195

$$[H]_{output}/[D]_{output} = 1.195[H]_{input}/[D]_{input} + 0.0414$$

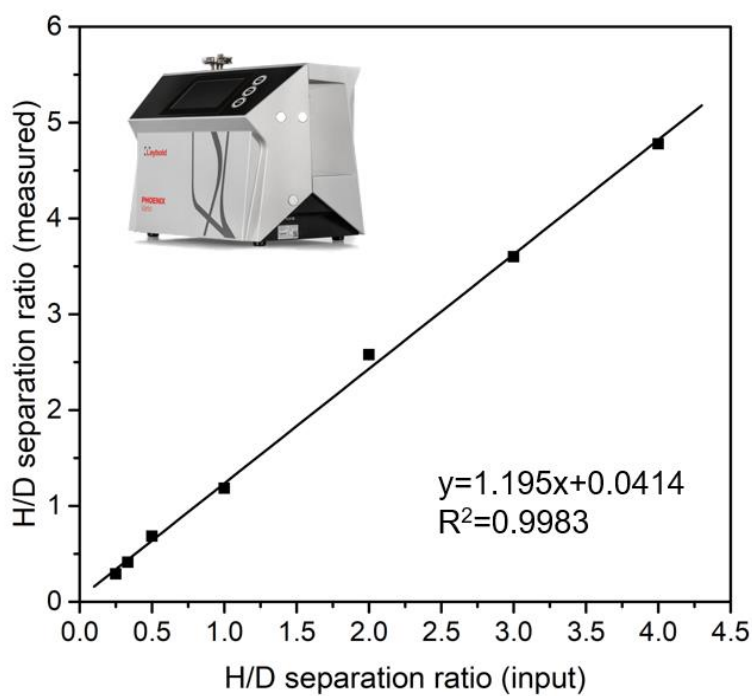

Fig S6. Calibration curve of helium ophthalmoscope